\documentclass{article}
\usepackage{amssymb}
\usepackage{amsfonts}
\newcommand{\Ref}[1]{(\ref{#1})}
\newcommand{\p}{\partial}

\newcommand{\eqa}{\begin{eqnarray}}
\newcommand{\neqa}{\end{eqnarray}}
\newcommand{\equ}{\begin{equation}}
\newcommand{\nequ}{\end{equation}}
\newcommand{\no}{\nonumber\\}
\def\w{\wedge}

\newcommand{\scr}{\rm\scriptscriptstyle}
\let\eps=\epsilon

\begin{document}

\title{{\bf On the perturbative expansion of a quantum field theory around a topological sector}}

\author{ {\bf{Carlo Rovelli${}^{ab}$\footnote{rovelli@cpt.univ-mrs.fr, simone.speziale@roma1.infn.it}\ , 
Simone Speziale${}^{a}$} } 
\\[1mm] 
\em\small{${}^a$Dipartimento di Fisica
dell'Universit\`a ``La Sapienza", and INFN Sez.\,Roma1, I-00185 Roma,
EU.}\\[-1mm] \em\small{${}^b$Centre de Physique
Th\'eorique de Luminy, Universit\'e de la M\'editerran\'ee, F-13288
Marseille, EU. }}
\date{\small\today}

\maketitle
\begin{abstract}
\noindent 
The idea of treating general relativistic theories in a perturbative expansion around a topological
theory  has been recently put forward in the quantum gravity literature.
Here we investigate the viability of this idea, by applying it to conventional Yang--Mills theory on flat spacetime. We find that the expansion around the topological theory coincides with the usual expansion around the abelian theory, though the equivalence is non--trivial.
In this context, the technique appears therefore to be viable, but not to bring particularly new insights.  Some implications for gravity are discussed.
\end{abstract}

\section{Introduction}
Conventional quantum field theory (QFT) treats interacting theories in a 
perturbative expansion around the quadratic term of the action.
This procedure can be applied to general relativity (GR) by expanding the metric
field around a fixed background, yielding the background--dependent
perturbative quantum gravity of Feynman, DeWitt, Veltman and many others.
The theory is non--renormalizable \cite{Sagnotti}, and the approach
makes sense only for computing low energy  gravitons scattering.
The question is open whether the non--renormalizability
is an intristic feature of the theory --and GR must be modified to
be a consistent theory--, or it has to do with the perturbative expansion chosen. 
In this case, GR could be non--perturbatively  renormalizable, or even finite 
\cite{carlo}.   In this case however,  a perturbative 
expansion,  different from the standard one, could still be well defined.  

An implementation of this idea has been recently proposed in \cite{laurent}, based
on the fact that GR can be written as a modified BF theory. 
BF is a topological field theory that can be quantized exactly and is finite, as
first showed by Witten in 3d \cite{Witten1}. The relevance of topological theories also for
4d quantum gravity is suspected since long \cite{carloarea}, and has been exploited 
in a number of ways.  The idea of \cite{laurent}, in particular, is to construct
quantum GR via a perturbative expansion around the topological theory (see also
 \cite{Witten1}).
This approach is radically new and raises a number of questions. In particular,
the unperturbed theory has no local degrees of freedom. Does it make sense to expand
a non--trivial QFT around a topological theory?
In this paper, we address this question by studying such expansion 
in a simpler case: Yang--Mills (YM) theory on a flat background. 

In fact, YM theory too can be written as a modified BF theory \cite{bfym},
sometimes called BFYM.  The conventional perturbative expansion of 
BFYM, where the zero'th order is the kinetic term, 
has been investigated in detail in the literature (see for instance
the last reference in \cite{bfym}). Here we show, however, that 
the coupling constant $g_0$ can be moved in such a way that the zero'th order is BF theory.
Therefore, the expansion in $g_0$ becomes an expansion around the 
topological theory, precisely as the one proposed for quantum gravity.
We thus investigate the perturbative expansion of the theory in this 
formulation.  In particular, we compute the propagator in a power series in $g_0$.

We find that the resulting expansion  is equal to the standard one.
In the non--abelian case, an interesting subtlety arises: the 
leading non--vanishing order is abelian, as expected, in spite
of the non--abelian nature of the zero'th order BF term.
We discuss the implications of this result for gravity  in the final section 3.

The possibility of expanding YM around BF 
opens the way to a homogeneous treatment of quantum gravity and YM theory. 
A proposal in this direction is under study \cite{noi}, and the present
work provides support for the viability of the approach.

\section{Expansion around BF of Yang--Mills theory}

Consider YM theory with gauge group $\cal G$, say SU$(N)$. 
We call $A_\mu^a$ the YM connection. The index $a$ is in the algebra $\mathfrak g$ of
$\cal G$. The YM curvature is $F_{\mu\nu}^a=\p_\mu A_\nu^a-
\p_\nu A^a_\mu + f^{abc}A_\mu^bA_\nu^c$, where $f^{abc}$ are the gauge group structure
constants. We use greek letters  $\mu=0, \ldots, 3$ for spacetime indices 
and $[\mu\nu]=\mu\nu-\nu\mu$.
For a compact notation, we use the formalism of $p$-forms, where 
$F= \frac{1}{2}F_{\mu\nu} dx^\mu\w dx^\nu=dA+\frac{1}{2}[A,A]$
and introduce the Hodge star $*$, defined by 
$*F_{\mu\nu}=\frac{1}{2} \eps_{\mu\nu}{}^{\rho\sigma}F_{\rho\sigma}$. 
We also introduce a $\mathfrak g$-valued 2-form $B^a$.

Consider the following two actions:
\begin{enumerate}
\item{the standard second--order YM action 
\equ\label{action1}
S_{\rm \scriptscriptstyle YM}[A^a_\mu]
= \frac{1}{4g_0{}^2}\int d^4x\  F^a_{\mu\nu}F_a^{\mu\nu}=
-\frac{1}{2g_0{}^2}\int \ {\rm Tr}\ F\w *F ;
\nequ
}\item{and the first order action 
\eqa\label{action2}
S_{\rm\scriptscriptstyle BFYM}[B_{\mu\nu}^a, A_\nu^a] &=& 
i \int \; {\rm Tr}\; B\wedge F - \frac{g_0{}^2}{2} \int \;
{\rm Tr}\;B\w *B.
\neqa
}\end{enumerate}
The traces Tr are over the algebra indices.
The equivalence of the two actions can be easily checked: the equation of motion
$$
0= \frac{\delta S_{\rm\scriptscriptstyle BFYM}}{\delta B_{\mu\nu}^a} = 
\frac{i}{2} \epsilon^{\mu\nu\rho\sigma} F_{\rho\sigma}^a
+ g_0{}^2  B_a^{\mu\nu}
$$
implies
\equ\label{B}
B_{\mu\nu}^a =  -\frac{i}{2g_0{}^2}\ \epsilon_{\mu\nu}{}^{\rho\sigma} F^a_{\rho\sigma}.
\nequ
Plugging \Ref{B} into \Ref{action2} and using 
$\epsilon^{\mu\nu\lambda\tau}\epsilon_{\mu\nu\rho\sigma}
= - 2 \delta^{\;\lambda\tau}_{[\rho\sigma]}$, 
we obtain \Ref{action1}.
Therefore, the two actions describe the same classical theory.

For $g_0=0$ the action \Ref{action2} reduces to the action for BF theory.
When using the action \Ref{action1}, one usually rescales the connection
$A\mapsto g_0 A$. In this way the coupling constant is moved
in front of the commutator term of the curvature,
$
F=dA+\frac{1}{2}g_0[A,A].
$ 
For $g_0=0$, thus, the scaled YM action reduces to the quadratic term.

To construct the quantum theory, we define the (Euclidean) $n$-point Green functions
\equ\label{Gym}
{\Gamma^{\rm\scriptscriptstyle YM}}^{a_1\ldots a_n}_{\mu_1\ldots \mu_n}[x_1 \ldots x_n; g_0]:= \frac{1}{Z}\;
\int {\cal D} A \; A_{\mu_1}^{a_1}(x_1)\ldots A^{a_n}_{\mu_n}(x_n) \; e^{-\frac{1}{4}\int 
{\rm Tr} \, \left(dA + g_0 [A, A] \right)^2 - S_{\rm gf}(A)},
\nequ
for the (rescaled) action \Ref{action1}, and
\equ\label{Gbfym} {\Gamma^{\rm\scriptscriptstyle{BFYM}}}^{a_1\ldots a_n}_{\mu_1\ldots \mu_n}[x_1 \ldots x_n; g_0] 
:=  \frac{1}{Z}\;
\int {\cal D} B \;  {\cal D} A \; A^{a_1}_{\mu_1}(x_1)\ldots A^{a_n}_{\mu_n}(x_n) \;
e^{-i\int {\rm Tr} \,B \wedge F+\frac{g_0{}^2}{2} \int{\rm Tr} \, B\w *B - S_{\rm gf}(B, A)},
\nequ
for the action \Ref{action2}.
Here $Z$ is the partition function, namely the functional integral without the field
insertions, and $S_{\rm gf}(A)$ and $S_{\rm gf}(B, A)$ are appropriate gauge--fixing terms.
In the following we use the Lorentz gauge.

Formally, these Green functions are equivalent:
integrating over the $B$ field in \Ref{Gbfym}, we indeed obtain
\equ
{\Gamma^{\rm\scriptscriptstyle YM}}^{a_1\ldots a_n}_{\mu_1\ldots \mu_n}[x_1 \ldots x_n; g_0]=
\frac{1}{g_0{}^{n}}{\Gamma^{\rm\scriptscriptstyle BFYM}}^{a_1\ldots a_n}_{\mu_1\ldots \mu_n}[x_1 \ldots x_n; g_0].
\nequ
The factor $g_0{}^{-n}$ comes from the rescaling used in \Ref{Gym}.
Since $n$-point Green functions are the basic building blocks of QFT, the 
equivalence above would guarantee that the two actions give rise to the same quantum theory.
However, the equivalence is only formal, until we can actually define and evaluate the integrals. 
In QFT, the conventional way of doing so is by expanding the functional integrals
in $g_0$ around a gaussian integral, that can be evaluated.
If the equivalence holds, then the two expansions should coincide, order by order. That is, 
(getting rid of indices and arguments)
\equ\label{pert}
\Gamma^{\rm\scriptscriptstyle YM}{}^{(0)}+
g_0 \;\Gamma^{\rm\scriptscriptstyle YM}{}^{(1)} + g_0{}^2 \;
\Gamma^{\rm\scriptscriptstyle YM}{}^{(2)} + \ldots=
\frac{1}{g_0{}^{n}} \; \left[\Gamma^{\rm\scriptscriptstyle BFYM}{}^{(0)}+
g_0{}^2\;\Gamma^{\rm\scriptscriptstyle BFYM}{}^{(1)} +
g_0{}^4\;\Gamma^{\rm\scriptscriptstyle BFYM}{}^{(2)}+\ldots\right].
\nequ
However, there are different reasons for doubting this equivalence,
and it is far from obvious how the two expansions could coincide. In particular,
the two zero'th orders are extremely different:  
\begin{itemize}
\item{In \Ref{Gym}, the zero'th order is a free abelian theory, namely a gaussian integral,
which depends on the metric and has lost track of the non--abelian structure,
\[
\Gamma^{\rm\scriptscriptstyle M}{}^{a_1\ldots a_n}_{\mu_1\ldots \mu_n}[x_1 \ldots x_n]:= \frac{1}{Z}\;
\int {\cal D} A \; A^{a_1}_{\mu_1}(x_1)\ldots A^{a_n}_{\mu_n}(x_n) \; e^{-\frac{1}{4}\int 
 {\rm Tr}\,\left(dA \right)^2 - S_{\rm gf}(A)}.
\]
}
\item{In \Ref{Gbfym}, the other way around, 
the zero'th order is not gaussian, the theory depends on the non--abelian structure
and has lost track of the metric of spacetime,
\[  \Gamma^{\rm\scr BF}{}^{a_1\ldots a_n}_{\mu_1\ldots \mu_n}[x_1 \ldots x_n] :=  \frac{1}{Z}\;
\int {\cal D} B \;  {\cal D} A \; A^{a_1}_{\mu_1}(x_1)\ldots A^{a_n}_{\mu_n}(x_n) \;
e^{-i\int {\rm Tr} \,B \wedge F - S_{\rm gf}(B, A)}.
\]}
\end{itemize}
How can starting points that are so different give rise to the same Green functions? 
Below we show that they do. The differences are harmless, and \Ref{pert} holds, order by order.  For simplicity, we focus on the 2-point Green function.

The perturbative expansion of \Ref{Gbfym} reads
\eqa\label{Gbfym1}
\Gamma^{\rm\scr BFYM}{}^{ab}_{\mu\nu}[x, y; g_0]&=& 
\frac{1}{g_0{}^2} \; \frac{1}{Z} \; \sum_{k=0}^\infty \frac{1}{k!} (\frac{g_0{}^2}{2})^k
\int {\cal D} B \;  {\cal D} A \;A^a_\mu(x) A^b_\nu(y) \; \left( \int {\rm Tr}\, B\w *B \right)^k \;
e^{-i\int {\rm Tr}\,B \wedge F - S_{\rm gf}(B, A)} \no &=&
\frac{1}{g_0{}^2}\;\Gamma^{\rm\scr BFYM}{}^{(0)}{}^{ab}_{\mu\nu}[x, y] + 
\Gamma^{\rm\scr BFYM}{}^{(1)}{}^{ab}_{\mu\nu}[x, y] + 
g_0{}^2 \; \Gamma^{\scr BFYM}{}^{(2)}{}^{ab}_{\mu\nu}[x, y] + \ldots
\neqa
For $k=0$ we have 
\equ
\Gamma^{\scr BFYM}{}^{(0)}{}^{ab}_{\mu\nu}[x, y]=
\frac{1}{Z} \;
\int {\cal D} B \;  {\cal D} A \;A^a_\mu(x) A^b_\nu(y) \; 
e^{-i\int {\rm Tr}\,B \wedge F - S_{\rm gf}(B, A)};
\nequ 
the gauge--fixing term for the $B$ field is needed to cancel the 
components of the $B$ fields, which are integrated over, but
do not enter the BF action because of the Bianchi identity on the 
curvature.
We perform the ${\cal D}B$ integral and obtain 
\equ\label{zero}
\Gamma^{\rm\scr BFYM}{}^{(0)}{}^{ab}_{\mu\nu}[x, y]=\frac{1}{Z}\;
\int {\cal D} A \;A^a_\mu(x) A^b_\nu(y) \; \delta(F) \; e^{-S_{\rm gf}(A)}\equiv 0,
\nequ
because in the Lorentz gauge $F=0$ implies $A=0$.
Therefore, the zero'th order does not contribute to the physics:
its correlators vanish.

The leading order of \Ref{pert} is then proved if we can show that, in the Lorentz gauge,  
\equ\label{bello}
\Gamma^{\rm\scr BFYM}{}^{(1)}{}^{ab}_{\mu\nu}[x, y]=
\Gamma^{\rm\scr M}{}^{ab}_{\mu\nu}[x, y]
=\int \frac{d^4p}{(2\pi)^4}\, \frac{e^{-i p\ (x-y)}}{p^2} \,
\Big(\eta_{\mu\nu} - \frac{p_\mu p_\nu}{p^2}\Big)\, \delta^{ab}.
\nequ 
To do so, we
consider separately the abelian and the non--abelian cases.

\subsection{Free propagator: abelian case}
In the abelian case, the curvature is simply given by the exterior derivative
of the connection, $F=dA$, and both the curvature and the $B$ field are
gauge--invariant. The gauge--fixing term only concernes the connection. To calculate
the propagator, we introduce a source $j^\mu(x)$ for the connection, and write
($j\cdot A$ stands for $j^\mu A_\mu$)
\eqa
\Gamma^{\rm\scr BFYM}{}^{(1)}{}_{\mu\nu}[x, y] &=&
\frac{1}{2} \int {\cal D} B \;  {\cal D} A \;A_\mu(x) A_\nu(y)  \int B\w *B \;
e^{-i\int B \w dA - S_{\rm gf}(A)}
=\no &=& - \frac{1}{2} \frac{\delta^2}{\delta j^\mu(x)\delta j^\nu(y)} \;
\int {\cal D} B \;  {\cal D} A \int B\w *B \;
e^{-i\int B \w dA - S_{\rm gf}(A)+ i \int j \cdot A}\Big|_{j=0}.
\neqa
We perform first the ${\cal D} A$ integral. Because of the Lorentz gauge--fixing
term, this integral selects only the transverse components of the source, 
$j_{\scr T}^\mu := (\delta^{\mu}_{\nu} - \frac{p^\mu p_\nu}{p^2}) j^\nu$. 
Integrating by parts in the action, we thus get
\eqa\label{pert1}
\Gamma^{\rm\scr BFYM}{}^{(1)}{}_{\mu\nu}[x, y] =
- \frac{1}{2} \frac{\delta^2}{\delta j^\mu(x)\delta j^\nu(y)} \;
\int {\cal D} B \int B\w *B \;\delta(*d B + j_{\scr T}).
\neqa
In order to evaluate the remaining integral, we have to find solutions of the equation $*dB=-j_{\scr T}$,
which in  Fourier--transformed components reads
$
\frac{1}{2}\epsilon^{\mu\nu\rho\sigma}p_\nu B_{\rho\sigma}(p)=-j_{\scr T}^\mu(p).
$
This can be easily solved in $B$, to give
$$
\epsilon^{\alpha\beta\rho\sigma} B_{\rho\sigma}(p)= \frac{2}{p^2}\ p^{[\alpha} j_{\scr T}^{\beta]}(p).
$$
It follows that
$$
\int {\cal D} B \int B\w *B \;\delta(*d B + j_{\scr T})
 = - 2 \int \frac{d^4p}{(2\pi)^4}\ \frac{1}{p^2}\ j_{\scr T}^\alpha(p)\ j^{\scr T}_\alpha(-p).
$$
Plugging this result in \Ref{pert1} it is straightforward to obtain
$$
\Gamma^{\rm\scr BFYM}{}^{(1)}{}_{\mu\nu}[x, y]= 
\int \frac{d^4p}{(2\pi)^4}\, \frac{e^{-i p\ (x-y)}}{p^2} \,
\Big(\eta_{\mu\nu} - \frac{p_\mu p_\nu}{p^2}\Big).
$$
It is easy to check that in this abelian case $\Gamma^{\rm\scr BFYM}{}^{(k)}$ for all $k>1$, so
that we actually verify \Ref{bello} exactly.

\subsection{Free propagator: non--abelian case}
In the non--abelian case, the situation is more subtle. As above, we introduce a source $j^\mu_a$
for the connection, but also a source $\eta^{\mu\nu}_a$ for the $B$ field. 
We can then write (here $\eta\cdot B$ stands for $\eta^{\mu\nu} B_{\mu\nu}$)
\eqa
&& \Gamma^{\rm\scr BFYM}{}^{(1)}{}^{ab}_{\mu\nu}[x, y] =
\frac{1}{2} \int {\cal D} B \;  {\cal D} A \;A^a_\mu(x) A^b_\nu(y) \int {\rm Tr} \, B\w *B \;
e^{-i\int {\rm Tr} \,B \w F - S_{\rm gf}(B, A)}=\no && = 
\frac{1}{8} \frac{\delta^2}{\delta j^\mu_a(x)\delta j^\nu_b(y)} \;
\int dz \frac{\delta}{\delta \eta_{\rho\sigma}^c(z)}\frac{\delta}{\delta \eta^{\rho\sigma}_c(z)}\;
\int {\cal D} B \;  {\cal D} A \; 
e^{-i\int {\rm Tr} \,B \w F - S_{\rm gf}(B, A)+ i \int {\rm Tr} \,j \cdot A 
+ i \int {\rm Tr} \,\eta \cdot B}\Big|_{j=\eta=0}.\nonumber
\neqa
Once again, the gauge--fixing term for $B$ is needed
to cancel the components which do not enter the BF action.
It is convenient this time to perform the ${\cal D}B$ integral first, obtaining
\eqa\label{1}
\Gamma^{\rm\scr BFYM}{}^{(1)}{}^{ab}_{\mu\nu}[x, y] &=&
\frac{1}{8}\frac{\delta^2}{\delta j^\mu_a(x)\delta j^\nu_b(y)}
\int dz \frac{\delta}{\delta \eta_{\rho\sigma}^c(z)}\frac{\delta}{\delta \eta^{\rho\sigma}_c(z)}
\int {\cal D} A \; \delta\Big( *F(A) - \eta \Big) \;
e^{- S_{\rm gf}(A)+ i \int {\rm Tr} \,j \cdot A}\Big|_{j=\eta=0} \no &=&
\frac{1}{8} \frac{\delta^2}{\delta j^\mu_a(x)\delta j^\nu_b(y)} \;
\int dz \frac{\delta}{\delta \eta_{\rho\sigma}^c(z)}\frac{\delta}{\delta \eta^{\rho\sigma}_c(z)}\;
e^{i \int {\rm Tr} \,j \cdot A_\eta}\Big|_{j=\eta=0}=
\no &=&
- \frac{1}{8} \int dz \, \frac{\delta A_\eta{}_\mu^a(x)}{\delta \eta^{\rho\sigma}_c(z)}
\, \frac{\delta A_\eta{}_\nu^b(y)}{\delta \eta_{\rho\sigma}^c(z)}
\Big|_{\eta=0}.
\neqa
Here we have introduced
the solution $A_\eta$ of the equation
\equ\label{Feta}
F_{\mu\nu}^a(A)=\eps_{\mu\nu\rho\sigma}\,\eta_a^{\rho\sigma}
\nequ
for the connection. The equation is non--linear, and we do not know the analitic
form of the solution.
However, as we see from \Ref{1}, only the linear dependence of $A_\eta$ on $\eta$ matters.
Let us therefore expand $A_\eta$ in a power series in $\eta$,
\equ\label{Aeta}
A_{\eta}{}^a_\mu(x) = \int dy \ G_1{}^{ab}_{\mu\nu\rho}(x, y) \ \eta^{\nu\rho}_b(y) + 
\int dy dz \ G_2{}^{abc}_{\mu\nu\rho\sigma\tau}(x, y, z) \ \eta^{\nu\rho}_b(y) \eta^{\sigma\tau}_c(z)+\ldots
\nequ
When we insert \Ref{Aeta} in \Ref{Feta}, the leading order, in Fourier components, is
$$
p_{[\mu} \, G_1{}_{\nu]\alpha\beta}^{ab}(p)\, \eta_b^{\alpha\beta}(p)
=\eps_{\mu\nu\rho\sigma}\,\eta_a^{\rho\sigma}(p).
$$
Remarkably, $G_1$ is only sensible to the abelian structure
of the theory.
In the Lorentz gauge we can solve this equation, obtaining
\equ\label{G1}
G_1{}_{\nu\alpha\beta}^{ab}(p)\,\eta_b^{\alpha\beta}(p)=\frac{1}{p^2}
\,\eps_{\mu\nu\rho\sigma} \, p^\mu\,\eta_a^{\rho\sigma}(p).
\nequ
Using this in \Ref{1} we immediately get
\eqa 
\Gamma^{\scr BFYM}{}^{(1)}{}^{ab}_{\mu\nu}[x, y] &=&
- \frac{1}{2} \int dz \,G_1{}^{ac}_{\mu\rho\sigma}(x, z) \,G_1{}^{bc}_{\nu\rho\sigma}(y, z)= \no &=& 
\int \frac{d^4p}{(2\pi)^4}\, \frac{e^{-i p\ (x-y)}}{p^2} \,
\Big(\eta_{\mu\nu} - \frac{p_\mu p_\nu}{p^2}\Big) \, \delta^{ab}.
\neqa
The non--abelian structure, yet present at the beginning of the calculation, plays no role in
the final result. 

Notice
how the non--abelian structure comes into play at the next order:
when we evaluate $\Gamma^{\scr BFYM}{}^{(2)}$, the term $\left(\int {\rm Tr} \, B\w *B \right)^2$ gives four functional
derivatives $\delta/\delta\eta$, which make the quadratic dependence of $A_\eta$ on
$\eta$ enter the final expression. But \Ref{Feta} at quadratic order in $\eta$
depends on the non--abelian structure, and we expect to
obtain the usual $g_0{}^2$ corrections to the free propagator.
We do not perform here the explicit calculations.

From the discussion above, we conclude that the perturbative expansion
around BF of YM theory on flat spacetime coincides with the conventional one.

\section{Consequences for gravity}
Two different formulations of GR as modified BF theories have been considered in the literature,
as starting point for the quantisation: 
\begin{enumerate}
\item{
The Plebanski action, 
\equ\label{aP}
S[B_{\mu\nu}^{IJ}, \omega_\mu^{IJ}, \mu_{IJKL}]= \frac{1}{16\pi G}\int {\rm Tr} \,B \w F(\omega) +
\int \mu_{IJKL}\, B^{IJ} \w B^{KL},
\nequ
used in \cite{Plebanski}.
The latin indices are in the group $SO(3,1)$ (or $SO(4)$ in the Riemannian version).
This has been used for the quantum gravity models described in \cite{Barrett}.
}\item{
The MacDowell--Mansouri action,
\equ\label{aM}
S[B_{\mu\nu}^{IJ}, \omega_\mu^{IJ}]= \int {\rm Tr} \,B \w F(\omega) -8 \pi G
\int \eps_{IJKLM}\,v^M\, B^{IJ} \w B^{KL}, 
\nequ
used in \cite{MacDowell, Smolin}.
The latin indices are in the group $SO(4,1)$ ($SO(5)$ in the Riemannian version).
This is a more recent approach to the problem, and its quantisation
has been considered in \cite{laurent}.}
\end{enumerate}
Both actions describe modified BF theories. However, there is a substantial difference between the two. 
The additional term in \Ref{aP} is a constraint, whereas the additional term in \Ref{aM}
is a genuine interaction. The $B$ field in \Ref{aP} is a fundamental
variable (and $\mu$ is a Lagrange multiplier), while the $B$ field in \Ref{aM} is only a
Lagrange multiplier. Notice also the different position of the Newton's constant $G$.

The perturbative treatment of the additional term appears more natural with the action \Ref{aM}.
In the rest of the discussion we focus on this approach. We assume   ${8\pi G}/{\hbar}\ll 1$, and
we discuss the perturbative expansion
\equ\label{pertGR}
Z_{\scr GR} = 
\sum_{n=0}^\infty \frac{1}{n!}(\frac{-i8\pi G}{\hbar})^n \int {\cal D} B \; {\cal D} \omega \;
\left( \int \eps_{IJKLM}\,v^M\, B^{IJ} \w B^{KL} \right)^n \; e^{\frac{i}{\hbar} \int {\rm Tr} \,B \w F(\omega)}
\nequ
of the partition function of quantum GR.

It is important to distinguish the two cases, of compact or non--compact spacetime.  
If spacetime is non--compact,
appropriate boundary conditions on the fields must be chosen, in order for the functional
integrals to be well--defined.
In the previous discussion of YM on flat spacetime, we implicitely chose vanishing boundary conditions
for the fields at spacial infinity. Once the boundary conditions
are chosen, the classical equations of motion select only one solution (up to gauge),
and the functional integral defining the Green functions is generically picked
around this solution, as discussed above. 
In this situation, we showed in the previous section that the perturbative expansion around BF
coincides with the conventional one.
If we treat the non--compact case in gravity, we have to provide spacial boundary conditions
for the gravitational field.
The natural choice here is asymptotic flatness: $g_{\mu\nu}\mapsto \eta_{\mu\nu}$ at infinity.
When using the action \Ref{aM}, this is a condition on the $B$ field.   Once this condition is fixed, we expect the quantum theory to be picked around a classical
solution (see the discussion in \cite{noialtri}).   In this setting, we expect our analogy with
YM theory to hold through: namely, \Ref{pertGR}
should give rise to the conventional (non--renormalizable) perturbation theory in terms of
gravitons. From this perspective, the ``expansion around the topological sector'' of the theory 
does not appear to be different from the conventional expansion. 

The situation is far less clear in the compact case, where the background field that minimizes the gaussian action, or is selected by the delta function, is not determined by asymptotic conditions. In particular, it has been recently argued that  the definition of the theory on a compact spacetime region with boundaries is more suitable to extract physical information from background independent theories \cite{carlo, boundary,  robert,  leonardo}. This is perhaps the situation where the proposal of \cite{laurent} appears to be more interesting.  We leave the discussion on this case open for further developments.


\begin{thebibliography}{10}

\bibitem{Sagnotti}
  M.~H.~Goroff and A.~Sagnotti,
  ``The Ultraviolet Behavior Of Einstein Gravity,''
  Nucl.\ Phys.\ B {\bf 266} (1986) 709.

\bibitem{carlo}
C.~Rovelli.  \newblock {\em Quantum Gravity}.  \newblock (Cambridge
University Press, Cambridge 2004.) 

\bibitem{laurent}
L.~Freidel and A.~Starodubtsev,
``Quantum gravity in terms of topological observables,''
arXiv:hep-th/0501191.

\bibitem{Witten1}
  E.~Witten,
  ``(2+1)-Dimensional Gravity As An Exactly Soluble System,''
  Nucl.\ Phys.\ B {\bf 311} (1988) 46.

\bibitem{carloarea}
C.~Rovelli,
``The Basis of the Ponzano-Regge-Turaev-Viro-Ooguri quantum gravity model in
the loop representation basis,''
Phys.\ Rev.\ D {\bf 48} (1993) 2702

\bibitem{bfym}
M.~B.~Halpern,
  ``Field Strength Formulation Of Quantum Chromodynamics,''
  Phys.\ Rev.\ D {\bf 16} (1977) 1798.

M.~Schaden, H.~Reinhardt, P.~A.~Amundsen and M.~J.~Lavelle,
  ``An Effective Action For Yang-Mills Field Strengths,''
  Nucl.\ Phys.\ B {\bf 339} (1990) 595.
  
A.~S.~Cattaneo, P.~Cotta-Ramusino, F.~Fucito, M.~Martellini, M.~Rinaldi, A.~Tanzini and M.~Zeni,
  ``Four-dimensional Yang-Mills theory as a deformation of topological BF
  theory,''
  Commun.\ Math.\ Phys.\  {\bf 197} (1998) 571
  [arXiv:hep-th/9705123].

\bibitem{noi}
L. Freidel, C. Rovelli and S. Speziale
``Perturbative expansion for 3d Yang--Mills theory coupled
to quantum gravity in the spinfoam formalism,''
to appear.

\bibitem{Plebanski}
J.~F. Plebanski,
  ``On the separation between Einsteinien substructure,''
  J.\ Math.\ Phys.\  {\bf 12} (1977) 2511.
\\
  R.~Capovilla, T.~Jacobson, J.~Dell and L.~Mason,
  ``Selfdual two forms and gravity,''
  Class.\ Quant.\ Grav.\  {\bf 8} (1991) 41.
\\
  R.~De Pietri and L.~Freidel,
  ``so(4) Plebanski Action and Relativistic Spin Foam Model,''
  Class.\ Quant.\ Grav.\  {\bf 16} (1999) 2187
  [arXiv:gr-qc/9804071].
\\
  M.~P.~Reisenberger,
  ``Classical Euclidean general relativity from *left-handed area =
  right-handed area*,''
  arXiv:gr-qc/9804061.

\bibitem{Barrett}
  J.~W.~Barrett and L.~Crane,
  ``Relativistic spin networks and quantum gravity,''
  J.\ Math.\ Phys.\  {\bf 39} (1998) 3296
  [arXiv:gr-qc/9709028].
\\
  A.~Perez,
  ``Spin foam quantization of SO(4) Plebanski's action,''
  Adv.\ Theor.\ Math.\ Phys.\  {\bf 5} (2002) 947
  [Erratum-ibid.\  {\bf 6} (2003) 593]
  [arXiv:gr-qc/0203058].

\bibitem{MacDowell}
S.~W.~MacDowell and F.~Mansouri,
  ``Unified Geometric Theory Of Gravity And Supergravity,''
  Phys.\ Rev.\ Lett.\  {\bf 38} (1977) 739
  [Erratum-ibid.\  {\bf 38} (1977) 1376].

\bibitem{Smolin}
  L.~Smolin and A.~Starodubtsev,
  ``General relativity with a topological phase: An action principle,''
  arXiv:hep-th/0311163.

\bibitem{noialtri}
F.~Mattei, C.~Rovelli, S.~Speziale and M.~Testa,
  ``From 3-geometry transition amplitudes to graviton states,''
  arXiv:gr-qc/0508007.

\bibitem{boundary}
F. Conrady, L. Doplicher, R. Oeckl, C. Rovelli, M. Testa
``Minkowski vacuum in background independent quantum gravity,''
\emph{Phys. Rev.}  \textbf{D69} (2004) 064019

\bibitem{robert}
  R.~Oeckl,
  ``A 'general boundary' formulation for quantum mechanics and quantum
  gravity,''
  Phys.\ Lett.\ B {\bf 575} (2003) 318
  [arXiv:hep-th/0306025].

\bibitem{leonardo}
  L.~Modesto and C.~Rovelli,
  ``Particle scattering in loop quantum gravity,''
  arXiv:gr-qc/0502036.

  
\end{thebibliography}
\end{document}